# The Evaluation of Circuit Selection Methods on Tor

Mohsen Imani[1], Mehrdad Amirabadi[1], and Matthew Wright[2]

[1]The University of Texas at Arlington, TX, USA
[2]Rochester Institute of Technology, NY, USA
mohsen.imani@mavs.uta.edu,mehrdad.amirabadi@mavs.uta.edu,matthew.wright@rit.edu

**Abstract.** Tor provides anonymity online by routing traffic through encrypted tunnels, called *circuits*, over paths of anonymizing relays. To enable users to connect to their selected destination servers without waiting for the circuit to be build, the Tor client maintains a few circuits at all times. Nevertheless, Tor is slower to use than directly connecting to the destination server. In this paper, we propose to have the Tor client measure the performance of the pre-built circuits and select the fastest circuits for users to send their traffic over. To this end, we define and evaluate nine metrics for selecting which pre-built circuit to use based on different combinations of circuit length, Round Trip Time (RTT), and congestion. We also explore the effect on performance of the number of pre-built circuits at the time of the selection. Through whole-network experiments in Shadow, we show that using circuit RTT with at least three pre-built circuits allows the Tor client to identify fast circuits and improves median time to first byte (TTFB) by 22% over Tor and 15% over congestion-aware routing, the state-of-the-art in Tor circuit selection. We evaluate the security of the proposed circuit selection mechanism against both a relay-level and a network-level adversary and find no loss of security compared with Tor.

## 1 Introduction

Tor provides anonymity for millions of users around the world by routing their traffic over paths selected from approximately 7,000 volunteer-run relays.[1] Tor effectively hides the user among all the users, so having more users and more traffic enhances anonymity for all [5, 10]. Unfortunately, Tor users often face large delays and long download times, which can discourage users and thereby reduce anonymity.

The client's traffic in Tor goes through an encrypted channel, called a *circuit*, which passes through a path of three relays: the *guard*, the *middle*, and the *exit*. The Tor client selects these relays randomly, weighted by their bandwidth in order to balance the load on the relays. At any moment, however, a selected relay or its ISP may be congested, or the Internet path between a pair of relays may be congested or otherwise slow. Further, since relays are spread out over the globe, some paths may cross multiple intercontinental hops. A number of researchers have investigated ways to address these issues with improved *path selection*, selecting the relays on the path for better performance [22, 6, 26, 7, 21]. While this can partially address the problem, it remains an issue that a path with good average performance might be performing poorly right now.

---
[1] https://metrics.torproject.org/, accessed April 2017

In this paper, we explore another approach to improving performance in Tor, based on the idea of *circuit selection*. When the user makes a request in Tor, such as for a webpage, the Tor client attaches the new stream to a circuit by opening a SOCKS connection. If no circuits are currently available for use, the client will build a new one, which involves conducting a cryptographic handshake with each of the three relays in turn. To reduce this delay, the client generally maintains multiple open circuits, which means that the client can attach the user's request to a circuit immediately. Tor currently does not use any performance criteria in selecting a circuit among the available circuits. There has been very little research focusing on improving Tor performance by modifying the circuit selection mechanism. To the best of our knowledge, Wang et al.'s work on Congestion-Aware Routing (CAR) [27] [2] is the only proposed approach, in which the client picks the best circuit among the available circuits based on a metric that they call the *circuit congestion time* (see §3 and §5.1 for details). Wacek et al. evaluated a number of path-selection algorithms as well as CAR and found that CAR had the best combination of performance and anonymity [26]. We thus compare our approach to CAR as the state-of-the-art.

**Contributions.** CAR shows good results compared with unmodified Tor, or *Vanilla Tor*, but there are other metrics that could be used to pick the circuit. In this paper, we examine other metrics that might be more effective, either by themselves or in combination. We first found that the number of available circuits in Tor is often small, between one and three circuits, such that picking the best circuit for performance does not have much effect in practice. As the number of available circuits increases, the chance of finding a fast and high-performing circuit should increase. To this end, for each circuit selection criteria we study, we evaluate the impact of more available circuits in terms of both performance and security.

We make the following contributions:

- We define nine circuit selection approaches using the geographical length, circuit delay, congestion, or a combination of these. We evaluate each of the approaches and compare them experimentally.
- We explore the performance benefit gained by increasing the number of available circuits at the time of circuit selection for each of proposed circuit selection approaches.

---

[2] Wang et al. introduced both circuit-selection and relay-selection mechanisms, but the relay-selection mechanism was found to have negligible impact on performance [27].

- We show the results of experiments on our approaches in Shadow, following the methodology of Jansen et al. [14]. We find using circuit RTT offers significant performance improvements over CAR, the current state-of-the-art in Tor performance research. In particular, our recommended approach provides a 15% reduction in median time to first byte and a 5% reduction in median time to last byte compared to CAR.
- We also measure the security of our recommended circuit selection approach with the rate of the compromised paths in the presence of relay-level and AS-level adversaries. We find that our approach provides anonymity in line with Tor.

## 2 Background

Tor is a volunteer-based overlay network providing anonymity online. More details are available at http://www.torproject.org/ and in the original design paper [11]. Here we provide a brief overview, touching mainly on details that are relevant to our work and eliding some details for clarity.

Relays in the Tor network are run by volunteers, who provide information about their donated bandwidth, IP address and ports, and exit policies—the addresses and ports they are willing to be connected to external Internet destinations—to a small group of servers called *directory authorities*. Directory authorities keep track of the relays' statuses and then mutually agree on the list of relays, called the *consensus*. The Tor client gets the information about the relays by downloading the consensus file from the directory authorities or their mirrors. Due the dynamics in the Tor network, the directory authorities publish the consensus documents hourly.

Then the client selects a three-hop path of relays. It first picks the last relay on the path, called *exit*. The exit is responsible to connect to the client's desired destination directly. Potential exits are limited to relays whose volunteers have set an *exit policy*, a specification of which ports (e.g. SSH, Web, file-sharing, SMTP) that the relay will allow to be connected to through it. Most relays have no exit policy, since being an exit exposes the volunteer to the liability of being directly associated with anonymous users' traffic. The client then picks the *entry*, which is limited to relays with greater reliability and higher bandwidth, and finally the *middle*, which can be any relay but is typically not an exit or entry. For all of these positions in the path, relay selection is done randomly with a bias towards higher bandwidth relays for load balancing and performance reasons. The bias is implemented, along with various other considerations, into a complex weighting function[3].

Once the client picks the relays for the path, it builds a circuit, which consists of three cryptographic tunnels, one between the client and each of the relays. The client first contacts the entry and uses an authenticated Diffie-Hellman handshake to share a secret key with it. The client and the entry use this secret key to encrypt and authenticate all of their communications, creating a secure tunnel between them. Through this tunnel, the client then asks the entry to extend the circuit to the middle. The client uses the same protocol to establish a secure tunnel with the middle that is layered inside of the tunnel between the client and the entry. Finally, the client uses the same protocol to extend the circuit to the exit through the tunnels to the entry and the middle. When the circuit is completed, then the client can attach a stream to it by opening a SOCKS connection. The stream's data is sent over the circuit and encrypted multiple times in a telescopic fashion such that no one on the path can link the source of the data to its destination.

---
[3] Full details at https://gitweb.torproject.org/torspec.git/tree/dir-spec.txt.

When the user, through her application, makes a request through Tor, the Tor client will first check to see if it has an open circuit available to attach the stream to. Typically, the client maintains one to three open circuits, as building a circuit takes time that would further slow down the user's experience. In particular, the client checks once per second to see whether there are at least two open circuits and creates new circuits if needed. Circuits that have been used for 10 minutes are marked as *dirty*, and they are not used for future connections, which means that a new circuit will be needed. More circuits can also be added if the user's application requires ports that are not allowed on the exit policies of the currently open circuits. Additionally, the Tor client maintains circuits for hidden services, which are servers that can only be accessed through the Tor network to protect the privacy of not only the user but the service itself, and one-hop circuits (to entries only) that are used to download the consensus. Geddes et al. report that the Tor client maintains an average of 10 circuits [13], though we note that usually only two of these are available for web browsing.

## 3  Related work

Researchers have addressed Tor performance issues in a variety of ways, such as modifying circuit scheduling [24], congestion control [27], traffic splitting [7], and incentives to encourage users to offer their bandwidth [12, 17]. In this section, we discuss the prior works on enhancing performance in Tor via improvements in circuit selection and relay selection.

Can et al. [24] propose a circuit scheduling mechanism that gives high priority to interactive traffic over bulk traffic on the same connection. This circuit selection mechanism has been deployed in Tor relays, but it has no impact on the client. Our circuit selection approaches are designed to improve the performance from the client side and are thus orthogonal to scheduling in the relays.

Wang et al. introduce *node latency* as a parameter to measure a relay's congestion [27]. In their approach, *congestion-aware routing (CAR)*, the client calculates *congestion time* as the difference between the most recently observed round-trip time (RTT) over the circuit and the minimum RTT for that circuit. The client measures the congestion time using both active and opportunistic methods and uses these measurements to avoid congested nodes during path selection (*long-term* congestion) and to avoid selecting congested circuits (*short-* *term* congestion). To avoid short-term congestion, they propose two methods: (i) choosing the best preemptively built circuit, which is a circuit selection mechanism, and (ii) switching to another circuit. To choose the best preemptively built circuit, they first randomly pick three circuits among all the preemptively built circuits, and then they select the circuit with the lowest congestion time. In their other method, switching to another circuit when the circuit is congested, they define a threshold of 0.5 seconds, and if the congestion time is more than this, they detach the stream from the circuit and attach it to another circuit. This causes a new TCP connection to the destination, which restarts the connection and harms the users' experience. To avoid *long-term* congestion, they propose a path selection mechanism in which the probability of selecting a relay is inversely proportional to its congestion time. Their results show improvement in quality of service and load balancing. In our evaluation of circuit selection methods, we also examine the use of congestion times as one of the possible metrics to use.

The current Tor client measures the Circuit Build Time (CBT), i.e. the time to construct the circuit, and uses this to discard slow circuits whose CBT is above a client-specific threshold. Annessi and Schmiedecker [8] propose that Tor should use the circuit round trip times (RTTs) in eliminating slow circuits instead of CBTs. In this method, the circuit RTT is actively measured after the circuit is built. If the RTT is longer than a threshold, then the circuit is marked as slow and no longer used. In their study, this provided only 3% improvement in the time to download the first byte of the destination website and offered mixed anonymity results. Our strategy in this paper is different from their approach. Rather than discarding slow circuits, we instead select the best performing circuit from a slightly larger set, and this offers better performance gains.

Several other works proposed modifications to Tor's path selection mechanism for improved performance. Snader and Borisov propose a modification to Tor path selection that helps Tor clients choose between anonymity and bandwidth depending on their needs [22]. They suggest using the distribution function $\frac{1-2^{s \times x}}{1-2^s}$ over the list of relays sorted by bandwidth, where $s$ can be raised for a greater chance of selecting high-bandwidth relays or lowered for a more uniform distribution. Sherr et al. use latency as a metric when selecting relays [21]. Their approach applies network coordinate systems so as to not require measurements between all pairs of nodes. Then clients can select paths with low overall latency. Akhoondi et al. [6] also aim to reduce latency, and they leverage geographical distance as a proxy for latency. In their scheme, LASTor, the relays are clustered geographically. They set up 15 geographically distributed DNS servers, and when the user makes a request, the client sends each of the servers a name resolution request. After receiving the IP addresses, the client locates the closest web server and then selects the lowest cost path through the clusters to the destination. Their results show 25% improvement over Tor.

Wacek et al. examine Tor path selection in a comprehensive study with experiments running many simultaneous clients [26]. They create a model of the Tor network to evaluate the recent published papers modifying path selection and show results for throughput, time to last byte (TTLB), and round-trip time (RTT). They tested Tor, Snader/Borisov [22], Unweighted Tor, in which Tor relays get selected uniformly at random, Coordinate [21] in which path selection is based on estimated pair-wise latencies, LASTor [6], and CAR [27]. Their investigation shows that path selection algorithms that do not consider bandwidth as a factor in relay selection have poor performance. They also showed that LASTor, in which each connection requires first making DNS resolution requests to 15 geographically distributed servers, showed poor performance. CAR had nearly the best performance in throughput and time-to-first-byte, plus it had anonymity approximately in line with Tor and significantly better than other high-performing algorithms. We thus select it for comparison in our work.

## 4 Model and Goals

### 4.1 Network Model

Testing new schemes on the live Tor network is technically challenging and could compromise users' anonymity or their harm their network performance [4]. Recently, tools such as Shadow [16] and ExperimenTor [9] have been developed to help model the Tor network and run experiments. We perform our simulations in Shadow [16, 3], a discrete-event simulator that runs the Tor code in a complete but scaled-down network. Shadow simulates the underlying network and it considers network attributes such as packet loss, bandwidth upstream and downstream, jitter, latency, and network edges. Jansen et al. provide a comparison [14] between doing experiments with Shadow, Tor itself, and ExperimenTor, another simulation platform.

For this study, we modified the Tor circuit selection mechanism to improve performance. All of our modifications are only on the Tor client and do not require any changes to the relays. In our performance evaluations, we used a scaled-down model of Tor, which consists of 1100 clients and 220 relays; this scaled-down model was built based on the procedures suggested by Jansen et al. [14] and measurements from the live Tor network (from July 2015).

### 4.2 Attacker Model

As with prior work in Tor performance [22, 21, 6, 26], our attention is more on performance characteristics than on attacks. We only seek to validate that our approach does not significantly weaken the anonymity provided by Tor currently. We evaluate the security of our proposed mechanisms in terms of both relay-level and network-level adversaries.

**Relay-Level Adversary Model.** In this adversary model, we assume that the adversary is running some Tor relays in the network with the goal of getting into the guard and exit positions of some circuits. An adversary in such a position can observe the entry and exit traffic and identify statistical correlations that link the clients to their destinations.

To evaluate the security of our proposed circuit selection mechanism, we simulate our proposed method along with CAR and *Vanilla* Tor in Shadow and randomly mark one of our guards and one of our exits as malicious relays. We then extract the streams and identify which ones were compromised. We repeat this process 10 times and measure the average compromise rates over the 10 runs.

**Network-Level Adversary Model.** The adversary can control some network components like ASes or IPXs. If the entry traffic and exit traffic of an anonymous connection traverse through the adversary's network components, the adversary can observe both sides of the traffic and de-anonymize the clients. We evaluate the security of our circuit selection mechanisms in the network level. We simulate the proposed mechanism in Shadow and extract the streams, including their paths. Because the proposed circuit method needs the traffic and RTTs, we can not use TorPS [18], the Tor path simulator, to generate the streams. TorPS is a tor path simulator that uses realistic network and user models, but it does not generate any traffic nor perform any encryption or decryption.

To determine the compromised streams, we use the algorithm proposed by Qiu and Gao [19] to infer the AS paths on both the entry side of the circuit (between clients and guards) and the exit side (between exits and servers). Qiu and Gao's algorithm exploits known paths from BGP tables to improve the inferred paths. In measuring the compromise rates, we consider the possibility of an asymmetric traffic correlation attack that can happen between the *data* path and the *ack* path, which is one of the RAPTOR attacks proposed by Sun et al. [23].

### 4.3 Design Goals

We seek an algorithm that meets the following goals:

1. Interactive use like web browsing should be significantly faster than Vanilla Tor and prior work.
2. Performance for bulk downloads should not be significantly slowed compared to Vanilla Tor.
3. Anonymity should be similar to what Vanilla Tor currently provides against our selected attack models.
4. We should avoid downloading large amounts of additional information from the directory servers.

We emphasize web traffic since delays in interactive use are more harmful to the user experience than delays in bulk downloads. We consider both response time, measured as time to first byte (TTFB), and total download time, measured as time to last byte (TTLB).

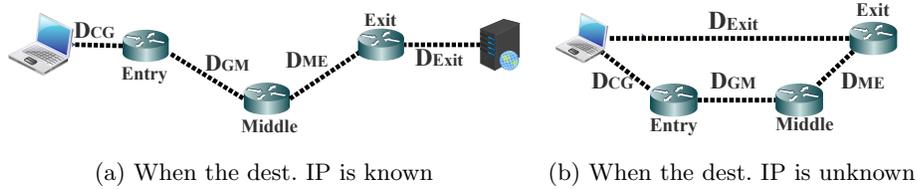

(a) When the dest. IP is known  (b) When the dest. IP is unknown

**Fig. 1.** Distances used to compute circuit length.

## 5 Circuit Selection

In this section, we explain how circuit selection can be improved by increasing the number of circuits and describe different possible performance criteria for use in circuit selection.

### 5.1 Performance in Circuit Selection

The Tor client does not use performance as a criteria when selecting from available circuits for attaching a stream. Wang et al. [27] propose to use the least congested circuit, but there are several possible performance characteristics to use instead. The number of available circuits are often small, such that existing circuit selection mechanisms are not effective in practice. Also, we know of no study testing the effect of changing number of circuits on performance-based selection.

To investigate the effect of circuit selection on Tor performance, we evaluate both the number of available circuits for the streams and the way to choose the best circuit among the available circuits. To set the number of circuits, we check once per second that there are at least $N$ circuits to support all recently used ports. If there are fewer than $N$ circuits, then we start building circuits to reach the threshold. We compare Vanilla Tor, which typically offers one or two circuits, with making at least $N = 3$ to $N = 5$ circuits available at all times. Given some number of circuits, we can then use various methods to select the best one. We compare various combinations of geographic circuit length, congestion, and round trip time (RTT).

**Metrics.** The three basic metrics that we use are geographic circuit length, congestion time, and round-trip time. To find the total geographic circuit length (or simply *length*) $L$ between the client and destination, we compute:

$$L = D_{CG} + D_{GM} + D_{ME} + D_{Exit} \qquad (1)$$

$D_{CG}$, $D_{GM}$, $D_{ME}$, and $D_{Exit}$ are shown in Figure 1a for when the destination IP address is known and in Figure 1b for when the IP address is not yet known. We use the opportunistic circuit measurements and latency model proposed by Wang et al. [27] to measure the circuit round-trip times and congestion times. Congestion time $T_c$ is measured as:

$$T_c = RTT - RTT_{min} \qquad (2)$$

where $RTT$ is the round-trip time and $RTT_{min}$ is the minimum RTT observed over that circuit. Wang et al. [27] showed that five measurements can effectively identify congested circuits. We thus measure and store the mean of the last five $T_c$ measurements as the congestion time of the circuit, and the mean of the last five RTTs as the circuit RTT.

### 5.2 Attaching Streams to Circuits

We consider nine different methods in handling streams using circuit length, congestion, and RTT.

1. *Congestion Only*: Pick the circuit with the lowest congestion time.
2. *Length Only*: Pick the shortest circuit.
3. *RTT Only*: Pick the circuit with the lowest RTT.
4. *Congestion then length*: Select the two lowest congestion times and pick the shorter circuit.
5. *RTT then length*: Select the two lowest RTTs and pick the shorter one.
6. *Length then congestion*: Select the two shortest circuits and pick the lower congestion time.
7. *Length then RTT*: Select the two shortest circuits and pick the lower RTT.
8. *RTT then Congestion*: Select the two circuits with the lowest RTTs and pick the lower congestion time.
9. *Congestion then RTT*: Select the two circuits with lowest congestion times and pick the lower RTT.

Since these circuit selection mechanisms are deterministic, given a set of candidate circuits, only one circuit from a set will be be used. These strategies will exploit the best circuit for the full 10-minute window that the circuit can be used. Since this means that the other circuits will go unused, we have the OP close any circuits that go unused for five minutes after their creation, leading to new circuits being opened. By itself, this might improve performance, as inferior circuits are closed in favor of untested circuits that may be better (or worse).

## 6 Circuit Selection Performance

We now evaluate the nine methods of selecting circuits for stream attachment and compare them with Tor and CAR.

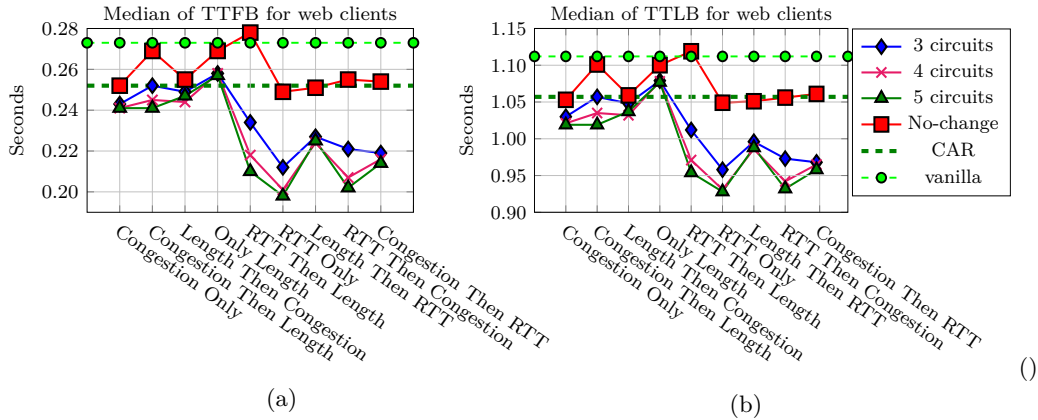

Fig. 2. **Circuit Selection:** TTLB and TTFB for web clients.

### 6.1 Network Configuration

We largely follow the experimental procedures suggested by Jansen et al. [14] and describe them here in brief. Shadow runs actual Tor code for accurate modeling; we used Tor version 0.2.5.12, modifying it as necessary to implement our methods and CAR. To generate a realistic Tor network topology, Shadow comes with topology generation tools that model a private Tor network based on a validated research study [14]. We used these tools and data from the Tor metrics portal to generate our private Tor network. Our Tor network includes 1100 clients, 220 Tor relays (52 exit relays, including exit-guard relays, and 49 guard relays), three directory authorities, and 220 HTTP destination servers.

Shadow uses an underlying topology that models the Internet. The default topology shipped with Shadow is very small, consisting of only 183 vertices and 17,000 edges, and is not a good representation of the Internet. For all the simulations in this paper we used the same Internet topology that was used by Jansen at al. [15]. This topology is provided by techniques from recent research in modeling Tor typologies [14, 18], traceroute data from CAIDA [2], data from the Tor Metrics Portal [25] and Alexa [1], and it includes 699,029 vertices and 1,338,590 edges. In our simulation, we tried different ratios of clients to relays, i.e. different congestion levels, and different average packet loss rates in the Internet topology and compared our results with Torperf data [25]. We found that a clients-to-relays ratio of 5:1 with 0.0025% packet loss provides us comparable results on TTFB and TTLB with Torperf data.

Our clients run Tor code in client-only mode and are distributed around the world in line with Tor usage statistics. We have two types of clients in our experiments: web clients and bulk clients. The 900 web clients download 320 KiB of data (the average page size [20]) and simulate web-surfing behavior by waiting between 1 to 20 seconds uniformly at random before starting the next download.

The 100 bulk clients download 5 MiB of data without pausing between the end of a download and starting the next one.

### 6.2 CAR: Congestion-Aware Routing

To compare our methods, we also simulated CAR, the circuit selection technique of Wang et al. [27]. They proposed opportunistic and active probing techniques to measure RTTs, which allows them to compute congestion times according to Equation 2, and they use these measurements to mitigate congestion using both an instant response for temporary congestion and a long-term response for low-bandwidth conditions. In our simulation, we follow the method of Wacek et al. [26], who also simulated CAR and ignored the long-term response due to its small impact on performance.

When attaching streams to circuits in CAR, we randomly select three circuits from the circuit list and pick the one that has the smallest mean congestion time from the five most recent measurements. If the mean of last five congestion times is more than 0.5 seconds for a circuit, we stop using the circuit for new streams.

### 6.3 Performance Results

Figure 2 shows the median time-to-first-byte (TTFB) and time-to-last byte (TTLB) for web clients. *Vanilla* represents unmodified Tor circuit selection, and *No change* represents the case where we do not modify the number of circuits from Tor, which typically has one or two circuits available at a time.

As shown in Figure 2, RTT is the best criterion to choose the circuit, with *RTT Only* as the best method overall. *RTT Only* has 15% lower TTFB than CAR (22% lower than *Vanilla*) for three circuits and 22% lower TTFB than CAR (27% lower than *Vanilla*) for five circuits. *RTT Only* also has 9% lower TTLB than CAR (13.8% lower than *Vanilla*) for three circuits and 12% lower TTLB than CAR (16% lower than *Vanilla* for five circuits. We speculate that RTT is the best criteria because it effectively captures both propagation delays and congestion time (queuing delays and transmission delays).

As expected, CAR is better than *Vanilla*, and *Congestion Only* with *no change* in the number of circuits performs the same as CAR, as both use the same criteria. Length turns out to be less effective compared to RTT or congestion times, particularly when the number of circuits is small. We note that *Length then RTT* performs fairly well for three or more circuits. Length may be suitable for gauging broad performance information, such as comparing a circuit with multiple intercontinental hops to one with no such hops, but poor at predicting the best circuit otherwise.

The TTFB and TTLB for *RTT Then Congestion* are slightly better than *Congestion Then RTT*, which indicates that RTT can narrow down candidate circuits better than congestion times. Results of both *RTT Then Congestion* and Congestion Then RTT are worse than *RTT Only*, which shows that mixing congestion time with RTTs will not provide better performance than using RTT by itself.

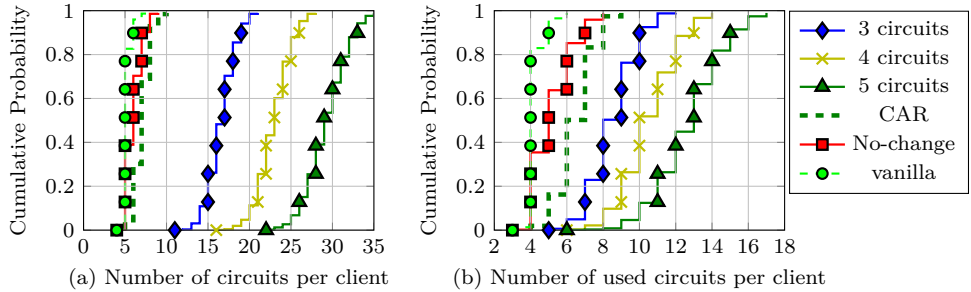

**Fig. 3.** CDF of the number of circuits created (a) and used (b) for web clients.

### 6.4 Circuit Creation Analysis

In our circuit selection strategies, we build more circuits than Tor's normal behavior, so it is important to understand the load this imposes on the network. Unfortunately, Shadow does not provide results regarding the load on nodes. To estimate changes in load, we compare the strategies based on the number of created circuits and used circuits. To see how many circuits our clients build, we simulate the circuit selection strategies in Shadow for one hour of simulated time, which leads to about 40 minutes of activity after 20 minutes of initialization. We extract the number of general-purpose circuits built by our web clients. Note that this does not including circuits built for hidden services or downloading the consensus, which is a consistent load across all schemes.

Figure 3 shows the CDF of created *general purpose* circuits and the CDF of used circuits, the circuits actually being used for transferring the data. We show results for web clients in *Vanilla*, CAR, and *RTT Only* with the same number of circuits as Tor, as well as *RTT Only* with $N = 3, 4, 5$ circuits. The median number of created circuits in *RTT Only* is 17, 23, and 29 circuits as we increase the number of circuits from three to five. *RTT Only* leads to building so many circuits due to proactively checking every second that there are $N$ circuits available for each recently used port, plus killing unused circuits after five minutes. Fig. 3.b shows how many of these created circuits have been used in transferring data. The median for used circuits in *vanilla*, CAR, and *RTT Only* with no change in the number of circuits is around four circuits, which means that they use all the circuits created and attach some stream to them. The median in *RTT Only* is 8, 10, and 13 circuits, respectively.

## 7 Security Analysis

In this section we examine the security of these circuit selection strategies, considering both relay-level and network-level adversaries. Our performance results show that *RTT Only* outperforms all the other circuit selection strategies and CAR. Therefore, in this section, we focus on the security analysis of *RTT Only*.

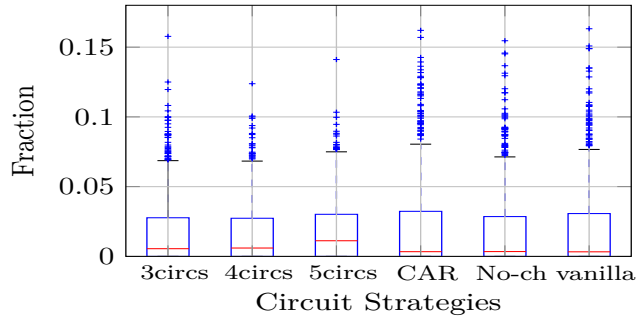

**Fig. 4.** *Relay-level adversary*: Distribution of compromise rates.

### 7.1 Relay-Level Adversary

In the relay-level adversary model, we assume that the adversary runs both guard and exit relays in the hope that his relays simultaneously occupy the guard and exit positions in some circuits. If the adversary can sit on the exit and guard position on a circuit, he can apply a traffic correlation attack and link the client to her destinations. These circuits and the streams attached to them are called compromised circuits and compromised streams, respectively. To analyze the security of the *RTT Only* strategy, we need to have access to RTTs (which include propagation delays, queuing delays, and transmission delays), which means we need to simulate a whole network. For this purpose, we again use Shadow to simulate the Tor network, and we use the same Tor network configuration as our performance evaluations in Section 6.1, which consists of 52 exit relays, including exit-guard relays, and 49 guard relays.

For the relay-level adversary, we randomly mark 10% of our guard bandwidth and 10% of our exit bandwidth as malicious guards and malicious exit relays in the network. Then we simulate CAR, *Vanilla*, and *RTT Only* with an unchanged number of circuits and then *RTT Only* using three to five circuits. We run 10 simulations, where the malicious guards and exits change in each run. 10 simulations for each case is reasonable considering that we have 52 exit relays, 49 guards, and simulations taking 11 hours.

Figure 4 shows box plots for the stream compromise rates for clients. The median of compromise streams is almost the same for *vanilla*, CAR, and *RTT Only* with *no change* in the number of circuits because the clients build almost the same number of circuits. As the number of circuits increases, the median of compromised streams rate increases due to the increase in circuits created by the clients. When the number of circuits increases, the chance of creating a circuit that has malicious relays on its guard and exit positions increases. As we see, *RTT Only* with five circuits has the highest compromised streams with a median of 1% while *vanilla* and CAR have a median compromised streams of 0.3%.

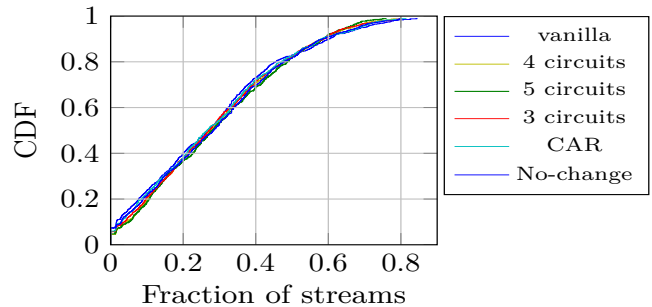

**Fig. 5.** *Network-level adversary*: CDF of compromise rates.

### 7.2 Network Level Adversary

In the network-level adversary model, we assume that the adversary controls an Autonomous System (AS). If the entry traffic (traffic between the client and guard) and exit traffic (traffic between the exit and server) traverse over a common AS, that AS can apply traffic correlation attack and link the client to her destinations.

To analyze the security of circuit selection strategies, we used our Shadow simulation results from the relay-level adversary for CAR, *Vanilla*, and *RTT Only* with an unchanged number of circuits and then *RTT Only* using three to five circuits. Because we did not add any relays to the network in evaluating the relay-level adversary model, as we only marked existing relays as malicious, we can re-use these results for this analysis. For each circuit selection approach, we extract all the generated streams by clients. The simulations generated approximately 730,000 streams for each circuit selection approach, or about 700 streams per client. For each stream, we used the algorithm proposed by Qiu and Gao [19] to infer the AS paths between clients and guards and between exits and servers. This algorithm exploits known paths from BGP tables to improve the accuracy of the inferred paths. In measuring the compromise rates, we consider the possibility of an asymmetric traffic correlation attack that can happen between the *data* path and *ack* path, which is one of the RAPTOR attacks proposed by Sun et al. [23].

Figure 5 shows the cumulative distribution of compromise rates for each strategy. As we see, when we increase the number of circuits from three to five, the the median compromise rate increases from 27.2% to 28.1% while the compromise rates of CAR and *vanilla* are 26% and 27%, respectively. CAR performs 1% better than *Vanilla*. These results show that using RTT and increasing the number of circuits even up to five circuits have a modest effect on the security against a network-level adversary.

## 8 Conclusions

Researches on Tor performance has mainly focused on modifying path selection, i.e. picking the relays for a circuit. In this paper, we improved Tor's performance by modifying the the circuit selection mechanism, i.e. picking from among already build circuits. Tor currently chooses circuits for transmitting user data without considering performance. We proposed nine different metrics for choosing high-performing circuits. We noticed that the number of pre-built circuits in Tor is very few, which makes the proposed metrics ineffective. We evaluated the impact of changing the number of pre-built circuits in conjunction with our proposed metrics on the performance. We found that using circuit RTT with only three pre-built circuits can improve the network responsiveness by 22% over Tor and 15% over CAR, the previous state-of-the-art scheme. Our security evaluations show that using circuit RTT with three pre-built circuits provides the same security as Tor in the presence of both relay-level and network-level adversaries. All of our proposed methods are implemented on the client side and they do not require any changes to Tor relays. Thus, they can be implemented easily in Tor.